\documentstyle[12pt]{article}
\newcommand{\lwig}{\mbox{\raisebox{.3ex}{$<$}$\!\!\!\!\!
$\raisebox{-.9ex}{$\sim$}}}
\newcommand{\gwig}{\mbox{\raisebox{.3ex}{$>$}$\!\!\!\!\!
$\raisebox{-.9ex}{$\sim$}}}

\newcommand{\sigpp }{ \mbox{$\sigma^{pp}_{{\rm multi-}w}       $} }

\begin{document}
\pagestyle{empty}
\title { Multiple $W_L$ Production From\\
Inelastic $W_LW_L$ Scattering At $\sqrt{\hat{s}} \gg M_H$ }
\pagestyle{empty}
\author{
D.A. Morris\thanks{morris@uclahep.bitnet},
R.D.~Peccei\thanks{peccei@uclahep.bitnet}
\\ {\it Department of Physics}
\\ {\it  University of California Los Angeles}
\\ {\it Los Angeles, California 90024-1547}
\and
R. Rosenfeld\thanks{rosenfeld@neuhep.hex.northeastern.edu}
\\ {\it Department of Physics}
\\ {\it Northeastern University}
\\ {\it Boston, MA 02115} }
\pagestyle{empty}
\date{UCLA Preprint 92/TEP/45 \\ November 1992}
\pagestyle{empty}

\maketitle
\pagestyle{empty}

\begin{abstract}
We explore the inelastic production of multiple longitudinal weak bosons
as a manifestation of a strongly interacting symmetry breaking sector. By
analogy with QCD, final states with large multiplicities are expected
to occur not far above the energy scale of the lowest resonances of
the underlying strong theory. We consider the feasibility of observing
such phenomena in the environment of a very high energy hadron collider.
\end{abstract}
\pagestyle{empty}
\newpage

\pagestyle{plain}
\section{Introduction}
\hspace*{\parindent}
One of the most interesting open questions in the Minimal Standard Model (MSM)
asks if the breakdown of $SU(2)_L \times U(1)_Y$ is caused by a weakly coupled
or strongly coupled theory. The MSM, with a Higgs potential
\begin{equation}
\label{eq:potential}
V = \lambda \left( \Phi^\dagger \Phi - \displaystyle{ v^2 \over 2 } \right)^2
\end{equation}
is prototypical of the first option, at least if $\lambda \ll 1.$
Models where the $SU(2)_L \times U(1)_Y$ breakdown is due to
fermionic condensates (e.g., $\langle \bar{T} T \rangle$) of some
underlying theory nicely typify the other possibility\cite{tcrefs}, since
condensate formation is symptomatic of strong coupling. These
two alternatives lead to quite distinct predictions for the scattering
amplitudes of longitudinal weak vector bosons $W_L \sim \{ W_L^\pm, Z_L\}.$
As is well known, through the equivalence theorem of
Cornwall, Levin and Tiktopolous\cite{clt,cg2}, high energy
$W_L W_L$ scattering is directly related to scattering of the corresponding
Goldstone bosons (denoted by
$w \sim \{ w^\pm,z \}$)
ensuing from the symmetry breakdown. If the symmetry
breaking sector is characterized by weak coupling, the scattering
among the $W_L$ will be weak. If, on the other hand,
there are strong interactions in the symmetry breaking sector,
these will be directly seen in $W_L W_L$ scattering.

This distinction between strong and weak coupling in the symmetry
breaking sector will only be apparent at high energies, since
at threshold the physics is the same. Let us momentarily focus
on the minimal Higgs model and rewrite the complex field $\Phi$ in
terms of the triplet of Goldstone boson fields $ w^\pm,z $ and the
physical Higgs field $H,$
\begin{equation}
\Phi =
 \left(
\begin{array}{c}
\displaystyle{ v + H + i z \over \sqrt{2} } \\
   \\
i w^-
\end{array} \right) .
\end{equation}
A simple calculation then leads to the following amplitude for
the scattering process $w^+ w^- \rightarrow z z,$
\begin{equation}
\label{eq:wpwmamp}
A( w^+ w^- \rightarrow z z )
= - 2 \lambda \left[
1 + \displaystyle{ 2 \lambda v^2 \over s - 2 \lambda v^2 }
\right],
\end{equation}
where $\sqrt{s}$ is the $w^+ w^-$ center of momentum system (c.m.s.) energy.
Analogous expressions are easily deduced for other channels.
At threshold, $s \ll M^2_H = 2 \lambda v^2, $ Eq.~\ref{eq:wpwmamp}
reduces to a simple expression which is {\it independent} of $\lambda,$
\begin{equation}
\label{eq:limit}
A( w^+ w^-  \rightarrow z z )
\longrightarrow
\displaystyle{s \over v^2 },
\end{equation}
reflecting that the dynamics at that point is solely determined
by the coset space of the breakdown --- here $O(4)/O(3) \sim
( SU(2)_L\times SU(2)_R ) / SU(2)_V$.
If the $SU(2)_L \times U(1)_Y$ breakdown is due to some strongly
coupled theory governed by the same global symmetry pattern as that of the
Higgs model, then here also $A( w^+w^- \rightarrow z z )$
at threshold would be given by Eq.~\ref{eq:limit}.

Above the threshold region, however, there are real distinctions
between the scattering amplitudes for, say, $w^+w^- \rightarrow z z$
predicted by the Higgs theory and that predicted by some model of
dynamical symmetry breakdown. The amplitude of Eq.~\ref{eq:wpwmamp} for the
Higgs case, when plotted as a function of $s,$ displays only one
remarkable feature --- a resonance pole in the $J=0$ channel at the
Higgs mass $M_H$. For $s \gg M^2_H$ this amplitude goes to a constant
\begin{equation}
A(w^+w^- \rightarrow z z ) \rightarrow - 2 \lambda
\end{equation}
which, for $\lambda \ll 1$ gives weak scattering.

For the case of dynamical symmetry breakdown, in complete analogy to
what happens in QCD for $\pi\pi$ scattering, one expects
significantly different behaviour. First of all, the scattering
amplitude should contain resonances in other partial waves besides
$J=0,$
\begin{equation}
A(w^+w^-\rightarrow z z) = 32 \pi \sum_J ( 2 J+ 1) P_J(\cos\theta)
a_J(s) .
\end{equation}
Secondly, at energies slightly above those where the first resonance forms,
one would also expect a rapid opening
up of inelastic channels. Because of this, the
partial waves $a_J(s)$ will not have unit strength
\begin{equation}
a_J(s) = \displaystyle{ 1\over 2 i }
\left[ \eta_J(s) e^{2 i \delta_J(s) } - 1 \right], \qquad \qquad
\eta_J(s) < 1.
\end{equation}

The search for strongly interacting effects in {\it elastic} $W_L W_L$
scattering at the Large Hadron Collider (LHC) and the
Superconducting Supercollider (SSC) has been a subject of intense interest in
recent years, resulting in literally hundreds of papers\cite{msc}.
On the other hand, {\it inelastic} $W_L W_L$ scattering,
after a pioneering paper by Chanowitz and Gaillard\cite{cg},
has hardly been investigated. (For recent speculations on the existence
of inelastic scattering involving additional Goldstone bosons
see Ref.~\cite{inelasticpapers}). This dichotomy in treatment
is not difficult to understand. Both the LHC and SSC
are machines which are barely above threshold, as far as $W_L W_L$ scattering
goes. Thus one is already pushed relatively hard to dig out a signal
for resonance formation in the $W_L W_L$ channel and it is essentially
hopeless to see any effects of the opening up of inelastic channels. However,
for a machine of even higher energy, like the proposed
ELOISATRON operating at
$\sqrt{s} = 200~{\rm TeV}$ these signals should become
more apparent. In some ways multi-$W_L$ production, when it is
prolific, may perhaps be a simpler signal to detect and it will
provide an equally distinct telltale sign of having strong dynamics
in the $W_L W_L$ channel. The purpose of this paper is to characterize
and quantify as best as one can this second aspect of having strong
dynamics in the symmetry breaking sector.

Because, as of yet, no one has a clear idea of the detailed dynamics
of a strongly coupled symmetry breaking sector, we will have to make
certain assumptions regarding the threshold for copious
multiparticle production and the nature of the signal
beyond that threshold. We will be guided in making these
assumptions by, among other things, the pattern of cross sections
and multiplicity distributions which are observed in hadronic
interactions. We will, however, make some further simplifying
assumptions whenever they appear warranted. This is a sensible
approach to take since there is no reason why an underlying
strong theory which produces the breakdown of
$SU(2)_L \times U(1)_Y$ should blindly copy QCD.
Furthermore, we also cannot pretend that the details we obtain
will be trustworthy. Nevertheless, we feel that the broad features
which emerge from our study --- despite our simplifying assumptions
--- should turn out to be generally correct.

The paper is organized as follows. In Section~\ref{se:characterization}
we review the connection between pions (denoted by
$\pi \sim \{\pi^\pm,\pi^0\}$ ) and the Goldstone bosons $w$
associated with weak symmetry breaking. Pursuing the notion
of strong dynamics in the weak symmetry breaking sector, we use
our inferred knowledge of strong inelastic
$\pi \pi$ scattering as a template and scale it up
from GeV energies to TeV energies to describe strong inelastic
$ww$ scattering. In Section~\ref{se:pp} we incorporate
the assumed strong $ww$ dynamics into calculations appropriate
for proton-proton scattering.
In Section~\ref{se:background} we present a rough
comparison of multi-$w$ signatures of strong inelastic $ww$ scattering
to multi-gauge-boson backgrounds expected in the MSM. In
Section~\ref{se:summary} we summarize our results and conclude.

\section{Scaling From GeV to TeV}
\label{se:characterization}
\hspace*{\parindent}
The reason for drawing an analogy between the Goldstone
bosons $w^\pm,z$ and pions
originates from the observation\cite{lqt,cg2} that the symmetry
breaking lagrangian of the MSM
is that of a $SU(2)_L \times SU(2)_R$ chirally symmetric linear sigma model
(LSM)\cite{gellmanlevy} --- the  same type of model which successfully
describes low-energy $\pi \pi$ scattering ($\sqrt{s_{\pi\pi}} ~\lwig~
1~{\rm GeV}$)\cite{weinberg}.
The correspondence between the two theories may be
expressed by associating
\begin{eqnarray}
 w & \rightarrow & \pi, \\
 H      & \rightarrow & \sigma   , \\
 v \simeq 246~{\rm GeV}     & \rightarrow & f_\pi \simeq 93~{\rm MeV} .
\end{eqnarray}
Thus, at least on a formal level, LSM
predictions for low-energy $\pi\pi$ scattering
may be related to MSM predictions for $w w$ scattering at a c.m.s. energy
$\sqrt{s_{w w}}$ by equating\cite{cg2,bunchofscalingreferences}
\begin{equation}
\label{eq:scale}
\sqrt{s_{ww}} = \displaystyle{ v \over f_\pi } \sqrt{ s_{\rm \pi \pi} }.
\end{equation}

In order to demonstrate the possible consequences of strong
inelastic $ww$ scattering we will, for definiteness, interpret
Eq.~\ref{eq:scale} literally and use it to map
inelastic $\pi \pi$ physics (inferred from known hadron phenomenology) to
hypothetically strong $ww$ interactions.
We can not overemphasize the extent to which this assumption
is largely unsubstantiated --- it is made in the spirit of simplicity
rather than absolute correctness. In effect, we will have elevated
the status of Eq.~\ref{eq:scale} from being a relationship between
the limiting cases of two {\it models} (the LSM of low energy pion physics
and the Higgs sector of the MSM) to assuming it is a relationship
between the {\it actual physics} of pions and the
{\it actual physics} of the Goldstone
bosons $w.$ Strictly speaking, the LSM description of elastic
$\pi \pi$ physics (and, by association, Eq.~\ref{eq:scale})
is only valid for $\sqrt{s_{\pi \pi}} ~\lwig~1~{\rm GeV}$
whereas we are interested in inelastic $\pi \pi$ physics typified
by $\sqrt{s_{\pi \pi}} ~\gwig~1~{\rm GeV}.$ With respect to the Goldstone
bosons $w$, the assumed literal equivalence to $\pi$
physics implies that if the MSM Higgs sector is only an effective theory
of some underlying theory like technicolor, then we have ignored
the possibility of their being additional Goldstone bosons in the
spectrum\cite{inelasticpapers}.
If we were to go beyond the MSM Higgs sector and assume, for example,
a $N_{\rm TC}$ technicolor model, we
should consider replacing Eq.~\ref{eq:scale} with
\cite{factorreference}
\begin{equation}
\label{eq:scale2}
\sqrt{s_{ww}} = \displaystyle{ v \over f_\pi }
\sqrt{\displaystyle{ 3 \over N_{\rm TC} } }
\sqrt{ s_{\rm \pi \pi} }.
\end{equation}
Again, in the interest of simplicity,
we shall ignore this embellishment and others\cite{golddugan}
except to note that in some instances
they suggest lower (and hence more accessible) energy scales
for $w w$ physics.

To the extent that we scale energies according
to a factor of $v/f_\pi$ and use $\pi \pi$ physics
as a guideline,  we now turn to parameterizing
multi-$w$ production with motivation from
hadronic phenomenology. A casual inspection of
data\cite{particledatabook} for
$\sigma^{p p}_{\rm total}(\sqrt{s}),$
$\sigma^{\pi p}_{\rm total}(\sqrt{s}),$
etc., reveals many salient features of hadron scattering.
The region $\sqrt{s}~\lwig~1-2~{\rm GeV}$
is typically dominated by resonance formation
characterized by large fluctuations in the cross section.
For $\sqrt{s}~\gwig~1-2~{\rm GeV}$ {\it elastic} cross sections
decrease rapidly, while total cross sections
remain almost constant apart from an eventual slow growth which does not
interest us here. This behaviour, corresponding
to the sudden onset of multiparticle production, originates from
the on-shell production and decay of many low-lying resonances.
Essentially all of the produced particles are pions.

Though direct $\pi\pi$ scattering is not experimentally feasible,
it is reasonable to expect that
$\sigma^{\pi \pi}_{\rm total}(\sqrt{s})$ also exhibits the
generic features of baryon-baryon and meson-baryon total cross sections.
In fact, since for fixed $\sqrt{s}$ a smaller fraction of the available
energy is invested in the rest masses of the initial pions
(compared to scattering involving baryons), one expects multipion
production to set in at even lower c.m.s. energies ({\it i.e.,}
with $\sqrt{s}$ closer to 1~GeV than 2~GeV). Scaling this up
by $v / f_\pi$ implies a corresponding onset of
multi-$w$ production for $ww$ c.m.s. energies above a threshold
of $\sqrt{\hat{s}_0} \simeq 2.5~{\rm TeV} - 5~{\rm TeV}.$
(The notation anticipates our use
of $\sqrt{\hat{s}_0}$ as a subprocess threshold
in proton-proton scattering.)
In reality, the multi-$w$ threshold would be determined by
the physics of the low-lying resonances of the strongly
interacting Higgs sector. Hence if no
resonances in $W_L W_L$ scattering are observed at the LHC or SSC,
then the corresponding scale of inelastic multi-$w$
production is pushed up (and
possibly out of reach of even the ELOISATRON).
In calculations we will use $\sqrt{\hat{s}_0} = 5~{\rm TeV}.$
In section~\ref{se:summary} we briefly discuss the possible, though
perhaps less plausible, scenario of a $\sqrt{\hat{s}_0}=1~{\rm TeV}$
threshold accessible to the SSC.

Treating $ w^\pm,z $ on the same basis,
we parameterize the multi-$w$ production cross section
at $w w$ subprocess c.m.s. energy $\sqrt{\hat{s}}$ by
\begin{equation}
\label{eq:subprocesscrosssection}
\hat{\sigma}^{ww}_{ {\rm multi}-w}
= \hat{\sigma}_0 \Theta ( \hat{s} - \hat{s}_0 ).
\end{equation}
The theta-function restricts our attention to
inelastic reactions and, in analogy with hadronic physics,
reflects the near constancy of the total cross section above
the inelastic threshold.
We can motivate a choice for $\hat{\sigma}_0$ in several ways.
On purely dimensional grounds, a constant total cross section of
$\hat{\sigma}_0 \simeq O(1/v^2) \simeq 6~{\rm nb}$
for strongly interacting Goldstone bosons is a reasonable guess.
A similar estimate follows
from scaling up $\sigma^{\pi \pi}_{\rm total}$
which, using the quark model additivity assumption\cite{kokedee},
is obtained from
\begin{equation}
\label{eq:ratio}
\sigma_{\rm total}^{\pi \pi}
=
\displaystyle{ \left( \sigma_{\rm total}^{\pi p} \right)^2
\over                 \sigma_{\rm total}^{  p p}            }
\simeq
\displaystyle{ \left( 25~{\rm mb} \right)^2
\over                 40~{\rm mb}           }
\simeq
 16~{\rm mb} ,
\end{equation}
where we have used data\cite{particledatabook}
for $\sigma_{\rm total}^{\pi p}$ and
$\sigma_{\rm total}^{p p}.$
Scaling up this value of $\sigma_{\rm total}^{\pi \pi}$
by $( f_\pi / v)^2$ gives $\hat{\sigma}_0\simeq 2~{\rm nb}.$
To be more conservative,  we will use $\hat{\sigma}_0 = 1~{\rm nb}$
for our subsequent numerical investigations.

We can estimate the multiplicities of Goldstone bosons in
inelastic $ww$ interactions by relating them to pion multiplicities
in $\pi \pi$ interactions. However, faced once more with an
absence of relevant $\pi \pi$ data, we must proceed indirectly.
We first connect the average charged pion
multiplicity in $\pi \pi$ interactions to
the average charged particle multiplicity
(essentially composed of $\pi^\pm$) measured in $e^+e^-$ annihilation.
A simple ansatz, motivated by studies which relate
mulitiplicity data from $pp$ collisions to multiplicity data
from $e^+e^-$ annihilation\cite{hrs}, is to assume that the average
charged pion multiplicity for $\pi \pi$ interactions at
c.m.s. energy $\sqrt{s_{\pi \pi}}$ is given by
the average charge multiplicity in $e^+ e^-$ annihilation
at c.m.s. energy
$\simeq \sqrt{s_{\pi \pi}} / 2.$ The factor of $1/2$ attempts to compensate
for the circumstance that not all of the energy $\sqrt{s_{\pi \pi}}$
is available for particle production.

As a second step, we connect the average $\pi^\pm$
multiplicities in $\pi \pi$ interactions to
$w^\pm$ multiplicities in $w w$ interactions by using
Eq.~\ref{eq:scale}.
For $w w$ interactions at a subprocess c.m.s.
energy $\sqrt{\hat{s}}$ we use the following
parameterization for the multiplicity of charged Goldstone
bosons ($\langle n_w \rangle = \langle n_{w^+} + n_{w^-} \rangle$),
\begin{equation}
\label{eq:avmult}
\langle n_w \rangle = \left\{
\begin{array}{ll}
3.32 - .408 \ln
\left( \displaystyle{ \hat{s}_{\rm eff} \over  1~{\rm GeV}^2  } \right)
+ .263 \ln^2
\left( \displaystyle{ \hat{s}_{\rm eff} \over  1~{\rm GeV}^2  } \right)
& \mbox{ if $\sqrt{\hat{s}_{\rm eff}} > 1.5~{\rm GeV} $} \\
 &  \\
3.2
& \mbox{ if $\sqrt{\hat{s}_{\rm eff}} \le 1.5~{\rm GeV} $}
\end{array} \right.
\end{equation}
where
\begin{equation}
\label{eq:relatecm}
\sqrt{ \hat{s}_{\rm eff} } =  \displaystyle{ 1 \over 2 }
           \displaystyle{ f_\pi \over v }    \sqrt{ \hat{s} } .
\end{equation}
The right hand side of
Eq.~\ref{eq:avmult} for $\sqrt{ \hat{s}_{\rm eff} } > 1.5~{\rm GeV}$
is a parameterization of data for the average charge multiplicity
in $e^+e^-$ annihilation\cite{delphi}.
Since the parameterization of Ref.~\cite{delphi} is not intended
to be used below 1.5~GeV (and is slightly pathologic in that
region) we take the average charge multiplicity to be
constant for $\sqrt{ \hat{s}_{\rm eff} } < 1.5~{\rm GeV}.$
As it turns out,
we will only make use of  Eq.~\ref{eq:avmult} in the
limited region
$.9~{\rm GeV}~\lwig~\sqrt{\hat{s}_{\rm eff}}~\lwig~2~{\rm GeV}$ hence
the precise details of the parameterization are largely irrelevant.
Furthermore, because $\langle n_w \rangle$ is a slowly
varying function of $\hat{s}_{\rm eff}$ in the region of interest,
the factor $1/2$ in Eq.~\ref{eq:relatecm} it is not of
major significance --- especially considering the more
speculative nature of the factor $f_\pi / v.$ Nevertheless
we include the factor of $1/2$ for completeness.

For $ww$ subprocess c.m.s. energies $\sqrt{\hat{s}} ~\gwig~5~{\rm TeV},$
Eq.~\ref{eq:avmult} gives $\langle n_w \rangle \simeq 3-4$ suggesting
that the distribution of $n_w$ about the average
should be well described by a Poisson distribution. Hence if we assume
\begin{equation}
\label{eq:neutral}
\langle        n_z \rangle                        =
\displaystyle{     \langle n_w \rangle \over 2       } ,
\end{equation}
(which is well motivated from hadronic physics) then the
probability of obtaining $n_w$ charged Goldstone bosons
and $n_z$ neutral Goldstone bosons becomes
\begin{equation}
\label{eq:poisson}
P(n_w, n_z, \sqrt{\hat{s}}) =
\displaystyle{ e^{-\langle n_w \rangle }
\langle n_w \rangle^{ n_w }
\over
n_w ! }
\displaystyle{ e^{-\langle n_z \rangle}
\langle n_z \rangle^{ n_z }
\over
n_z ! }.
\end{equation}
We will incorporate
Eqs.~\ref{eq:avmult}-\ref{eq:poisson} into
our quantitative analysis. Strictly speaking eq.~\ref{eq:poisson}
cannot be correct for $n_w + n_z = 0$ or $1$ because of
energy-momentum conservation. Since our interest will be
in high multiplicity states, ignoring this inconsistency actually
leads to more conservative results ({\it i.e.,} it will
lead to slightly smaller probabilities for high-multiplicity
final states).

\section{pp Cross Sections}
\label{se:pp}
\hspace*{\parindent}
Consider an inelastic $W_L W_L$ subprocess above a
threshold $\sqrt{\hat{s}_0} = 5~{\rm TeV}$
in the environment of a proton-proton ($pp$) collider.
In the c.m.s. of the hard subprocess
each initial $W_L$ has an energy $E_{W_L}$ such that
\begin{equation}
  \displaystyle{ E_{W_L}          \over  M_W }
\, = \, \displaystyle{ \sqrt{ \hat{s} } \over 2  M_W }
\,~\gwig~\, 30 .
\end{equation}
By the equivalence theorem\cite{clt,cg2}
we are then quite justified in replacing the
longitudinal gauge bosons $W_L^\pm,Z_L$
with the corresponding Goldstone bosons $w^\pm,z.$
With this equivalence in mind
we subsequently phrase many of our results
directly in terms of Goldstone bosons.

Let $\sigpp$ denote the $pp$ cross section for
multi-$w$ production.
The effective vector boson approximation\cite{evbaref} and
the parameterization of Eq.~\ref{eq:subprocesscrosssection} give
\begin{equation}
\label{eq:sigppeq}
\sigpp(\sqrt{s}) =
\sum_{w_i w_j}
\displaystyle{ 1 \over 1 + \delta_{ij} }
\int \, dx_1 \, dx_2 \,
f_{w_i}(x_1)  f_{w_j}(x_2)
\hat{\sigma}_0 \Theta( x_1 x_2 s - \hat{s}_0 ) .
\end{equation}
The double sum extends over $w_i \sim \{ w^\pm,z \}$
where $f_{w_i}(x)$ is the distribution function of
$w_i$ carrying a fraction $x$ of the original
proton momentum.

Specifically,
\begin{equation}
\label{eq:bosondistribution}
f_{w_i}(x) = \sum_k \int^1_x \displaystyle{ dy \over y }
f_k(y) P_{w_i / k}\left( \displaystyle{ x \over y } \right)
\end{equation}
where $f_k(y)$ is the distribution function for quarks (or anti-quarks)
of species $k$ inside a proton.
The splitting function $P_{w_i / k }(x)$
is the probability that a Goldstone boson $w_i$
(or, more appropriately, the associated longitudinal gauge boson) carries
away a momentum fraction $x$ from a parent quark of species $k.$
Since we are interested in $w w$ subprocess energies
$\sqrt{\hat{s}} \gg M_W$ we are justified in using the
leading logarithmic form of the effective vector boson approximation
for the splitting functions found in the literature\cite{evbaref}.
In our calculations we employ the Morfin-Tung SL-fit
leading order distribution functions $f_k(y)$ from Ref.~\cite{mfdist}
evaluated at $Q^2 = M_W^2$ (which is the
scale implied by the emission of an on-mass-shell longitudinal boson).

For later convenience we rewrite Eq.~\ref{eq:sigppeq} in the form
\begin{equation}
\sigpp(\sqrt{s}) =
\int_{ \hat{s}_0 / s }^1 \, d\tau \, {\cal L}(\tau) \, \hat{\sigma}_0 ,
\end{equation}
where ${\cal L}(\tau)$ is the $w w$ luminosity function
\begin{equation}
\label{eq:luminosity}
{\cal L}(\tau) =
\sum_{ij} \displaystyle{ 1 \over 1 + \delta_{ij} }
\int_\tau^1 \, \displaystyle{ d x_1 \over x_1 } \,
f_{w_i}(x_1)  f_{w_j}\left( \displaystyle{ \tau \over x_1 } \right).
\end{equation}
Figure~1 plots $\sigpp / \hat{\sigma}_0$ as a function
of $\hat{s}_0 / s.$
With $\sqrt{\hat{s}_0} = 5~{\rm TeV},$ $\hat{\sigma}_0 = 1~{\rm nb},$
the ELOISATRON ($\sqrt{s} = 200~{\rm TeV}$)
gives $\sigpp \simeq 190~{\rm fb}.$ For purposes of
comparison, a machine luminosity of
$10^{33}~{\rm cm}^{-2}{\rm s}^{-1}$ over a nominal $10^7~{\rm s}$
year gives an integrated luminosity of $10~{\rm fb}^{-1}$ ---
corresponding to 1900~events (before considering branching fractions,
detector acceptance, efficiency etc.).

The Goldstone boson multiplicity distribution
for $pp$ collisions may be
expressed in terms of the $ww$ luminosity of Eq.~\ref{eq:luminosity}
and the subprocess multiplicity of Eq.~\ref{eq:poisson} ,
\begin{equation}
\label{eq:exclusivemult}
\tilde{P}(n_w,n_w,\sqrt{s} )
=
\displaystyle{
  \int_{ \hat{s}_0 / s }^1  \, d\tau \, {\cal L}(\tau)
  P( n_w, n_z, \sqrt{\tau s} )
\over
  \int_{ \hat{s}_0 / s }^1  \, d\tau  \, {\cal L}(\tau)
} .
\end{equation}
Figure~2 shows the total multiplicity distribution
for Goldstone bosons
\begin{equation}
\label{eq:ptilde}
\tilde{P}(n,\sqrt{s}) = \sum_{n_z,n_w}
\tilde{P}(n_w,n_z,\sqrt{s}) \, \delta_{n~n_z+n_w}
\end{equation}
for $\sqrt{s} = 200~{\rm TeV}$ and $\sqrt{\hat{s}_0}=5~{\rm TeV}.$
Because the luminosity ${\cal L}(\tau)$ falls so rapidly,
most interactions occur just
above the subprocess threshold $\sqrt{\hat{s}_0};$
for our purposes, we could equally well have used
$P(n_w,n_z,\sqrt{\hat{s}_0})$ instead of
$P(n_w,n_z,\sqrt{\tau s})$ in Eq.~\ref{eq:exclusivemult}.
In other words, the high-multiplicity tail in Fig.~2 is determined
largely by fluctuations in the multiplicity for subprocesses
just above threshold and not by fluctuations in the subprocess energy
itself.

Turning to the kinematics of multi-$w$ production, we
again get inspiration from hadron physics.
In analogy with QCD where the bulk of multiparticle production
is characterized by limited transverse momentum
$\langle p_T \rangle_{\rm QCD} \simeq 400~{\rm MeV} ~(\simeq
O(\Lambda_{\rm QCD}) \simeq O(f_\pi) )$
it is natural to assume that multi-$w$ production
is similarly governed by a parameter $\langle p_T \rangle_w$.
Two plausible $O(v)$ guesses are
\begin{equation}
\label{eq:smallpt}
\langle p_T \rangle_{w}  \simeq  M_W ,
\end{equation}
and
\begin{equation}
\label{eq:bigpt}
\langle p_T \rangle_{w} \simeq
\displaystyle{ v \over f_\pi } \times
\langle p_T \rangle_{\rm QCD} \simeq 1~{\rm TeV}.
\end{equation}

A simple choice for the subprocess cross section
$ \hat{\sigma}( ww \rightarrow n~w ) $ reflecting limited $p_T$ is given by
\begin{eqnarray}
\label{eq:phasespace}
\lefteqn{ d\hat{\sigma}( w w \rightarrow n~w) =} \\
& & F_n(\hat{s},\hat{s}_0,M_W)
 \delta^4 \left( P_{\rm total} - \sum_i p_i \right)
\prod_{i}
\displaystyle{ d^3 p_i \over 2 E_i }
\exp \left[ - \displaystyle{ (p^2_T)_i \over
 2 \langle p_T \rangle_{w}^2 } \right] ,  \nonumber
\end{eqnarray}
where $P_{\rm total}$ is the total four-momentum of the system and
$p_i$ are the individual final state momenta. The
normalization $F_n(\hat{s},\hat{s}_0,M_W)$ is chosen
so that after integrating over phase space and summing over all
possible multiplicities one reproduces the total $ww$
cross section of Eq.~\ref{eq:subprocesscrosssection}.

Figures~3,4 show the laboratory distributions for gauge boson
rapidity $y_{w}$ and $p_T$ for the case
$n=8$ for a subprocess energy $\sqrt{\hat{s}} = 5~{\rm TeV}$ at
$\sqrt{s} = 200~{\rm TeV}.$
A cursory inspection of the $p_T$ distribution for the
scale choice $\langle p_T \rangle_{w} \simeq 1~{\rm TeV}$
confirms what could have been trivially anticipated: that
damping transverse momentum beyond $\langle p_T \rangle_{w}$
is irrelevant if the average subprocess c.m.s. energy per particle
$\sqrt{\hat{s}} / n \ll \langle p_T \rangle_{w}$ ---
for the example at hand $\sqrt{\hat{s}} / n \simeq 625~{\rm GeV}$
is already smaller than $\langle p_T \rangle_w = 1~{\rm TeV}$ so that
the eight bodies are effectively distributed according to pure
phase space. Consequently, if the $p_T$ scale relevant to
strong $w$ dynamics is indeed $\sim 1~{\rm TeV},$
multi-$w$ events at $\sqrt{s}= 200~{\rm TeV}$ are essentially
spherical and would easily
be contained in a laboratory detector. Moreover, the
large $p_T$ involved (characterized by the minimum of
$\sqrt{\hat{s}} / n$ and $\langle p_T \rangle_{w}$)  gives
rise to unambiguous high-$p_T$ leptons and jets which have no
simple standard model background (at least for the production of
very many bosons).

The implications of the alternative scale
$\langle p_T \rangle_{w} = M_W$ are less dramatic but
deserve closer attention. Though the corresponding laboratory
rapidity distribution is certainly broader,  even on an
event by event basis, an overwhelming
majority of the Goldstone bosons
in the signal still fall within $|y| \leq 3 $ relevant
to a realistic detector. Of more concern is the $p_T$
distribution since the standard model background production of $W$ bosons
or $t$ quarks which subsequently decay to $W$ bosons is also characterized
by $p_T$ of $O(M_W)$ or $O(m_t).$  We will return to such backgrounds in
the next section.

\newpage
\section{Signatures and Backgrounds}
\label{se:background}
\hspace*{\parindent}

The speculative nature of inelastic multi-$w$ production makes
the issue of backgrounds difficult to assess in a completely
satisfactory manner. Nevertheless, by introducing a few rough,
yet plausible, assumptions we can assess the feasibility of
observing a multi-$w$ signal against a background of generic
strong and electroweak processes. In this section we decompose
the $pp$ multi-$w$ cross section in terms of the experimentally
more relevant cross sections for jet plus lepton
signatures and compare them to MSM background processes.
Since we will only concentrate on rather broad features of the
signal and background, we will not attribute overdue significance
to our quantitative results; we are only interested
in the plausibility of observing inelastic multi-$w$ production.

For definiteness, we will restrict our
attention to $pp$ collisions at
$\sqrt{s} = 200~{\rm TeV}$ where $\hat{\sigma}_0 = 1{\rm~nb}$,
$\sqrt{ \hat{s}_0 } = 5 {\rm~TeV}$
corresponds to $\sigpp \simeq 190 {\rm~fb}.$
Using the $n_w,n_z$ multiplicity distribution of
Eq.~\ref{eq:exclusivemult} and the known decay branching fractions for
$W$ and $Z$ bosons (see Table~1),
it is a straightforward exercise to calculate the signal cross
section for all possible lepton and jet signatures.
It is convenient to characterize signatures by the simultaneous
specification of $( n_{Z \rightarrow \ell \bar{\ell} },
   n_{W \rightarrow \ell \nu       },
   n_{\rm jets} ) $ where (1)
$ n_{Z \rightarrow \ell \bar{\ell} }$ is the number of
Z decays to $e$ or $\mu$ pairs (which we assume are detected and
reconstructed with $100\%$ efficiency),
(2)  $n_{W \rightarrow \ell \nu       }$ is the number
of $e$ or $\mu$ presumably
arising from leptonic $W$ decays and
(3) $n_{\rm jets}$ is the number of jets presumably arising
from the hadronic decays of $W$ and $Z$ bosons.
As as simplification, we will assume that all leptons and jets
from signal processes (and eventually, all background processes as well)
are individually resolved and meet minimum acceptance requirements.
In view of the kinematics attributed to multi-$w$ production,
we could set minimum acceptance requirements in the
neighbourhood of $|y| \le 3$ and $p_T~\gwig~40~{\rm GeV}$ and be
confident that essentially all signal events fall with a realistic
detector.

Before we present signal cross sections for
various signatures, let us consider possible backgrounds to multi-$w$
production. While the prospect of multiple gauge bosons exploding
into existence conjures up images of spectacularly rich multi-jet plus
multi-lepton signatures, it turns out that at
$\sqrt{s}=200~{\rm TeV}$ such configurations
will be commonplace from much less speculative processes. For example,
unless one can reliably distinguish between longitudinal and transverse
gauge bosons, there will be a large background of
multiple W boson events originating from the decays of
copiously produced $t$ quarks. In Ref.~\cite{bargerweakbosonpaper}
Barger~{\it et al.} have considered the possibility of observing
multiple gauge bosons at the LHC and SSC from $t$ quark decay, $H$
boson decay and generic electroweak production processes; their results will
provide a convenient starting point for our background estimates.

By minimally extrapolating Fig.~9 of Ref.~\cite{bargerweakbosonpaper}
we obtain the MSM cross sections of Table~2 (for
$m_t = 140~{\rm GeV}$ and $m_H = 400~{\rm GeV})$ relevant
to the direct and indirect ({\it i.e.}, coming from the decays of $t$ or $H$)
production of multiple gauge bosons at $\sqrt{s} = 200~{\rm TeV}.$
Not all of the cross sections of Table~2 will be serious backgrounds
--- most are included so that we can assess their significance relative to
$\sigpp~\simeq 190~{\rm fb}.$ Furthermore, since we are working on the
premise of a strongly coupled Higgs sector (which generally implies a
heavy Higgs boson), the cross sections of Table~2 involving Higgs bosons
are most likely overestimates since they assume a relatively light
$m_H = 400~{\rm GeV}.$

With the exceptions of $\sigma(t \bar{t})$ (which
is from an $O(\alpha_s^3)$ calculation) and
$\sigma(t\bar{t}b\bar{b}),$ none of the processes of Table~2
include QCD corrections to weak boson production.
Additional backgrounds may be obtained
by ``dressing'' each process with QCD radiation.
For example, in addition to $t  \bar{t} Z$ production
we should also consider the production of
$ t \bar{t} Z g, $ $t \bar{t} Z g g,$ etc.,
corresponding to final states with additional
hadronic jets. We will return to these additional backgrounds
after we consider those of Table~2.

Assuming identical acceptance criteria as for the signal, it
is a simple combinatoric exercise to convert the cross sections
of Table~2 into background cross sections for
various $( n_{Z \rightarrow \ell \bar{\ell} },
n_{W \rightarrow \ell \nu       },
n_{\rm jets} ) $ signatures. For both our signal and
background calculations we used the branching fractions of
Table~1. The only nontrivial aspect of
the background calculation is the treatment of $b$-quark decays. Since
$b$ quarks from $t$ decay retain a non-negligible fraction of the
original $t$ quark transverse momentum, $(p_T)_t \simeq O(m_t/2)$,
high $p_T$ electrons or muons from leptonic $b$ decays are a background
to the leptonic decays of on-shell W bosons.
In an idealized decay $b \rightarrow \ell + {\rm jet},$
kinematics dictates that the laboratory
angular separation of the lepton and jet
decreases as the $p_T$ of the parent $b$ quark increases.
Following Barger~{\it et al.}\cite{bargerweakbosonpaper}
we assume that we can exploit this small separation
and introduce an effective rejection factor
of $\simeq 1/40$ for high $p_T$ leptons from $b$ decays.
Operationally this means that we suppress cross sections
from final states containing $n$ leptons from $b$
decays by a factor of  $(1/40)^n.$

Tables~3,4 list the signal to background ratios for signatures
corresponding to $ n_{ Z \rightarrow \ell \bar{\ell} } = 0, 1$ respectively.
The total signal cross section for ``gold plated'' signatures with
two or more leptonically reconstructed $Z$ decays is negligible.
For each table entry the quantity in parenthesis is the background
summed over the contributing processes of Table~2.
As mentioned above, additional backgrounds arise if one
considers dressing the processes of Table~2 with QCD radiation.
A crude way of accounting for these additional backgrounds is to
assume, for example, that
$\sigma(t \bar{t} Z + {\rm jet} ) \simeq \alpha_s \sigma(t \bar{t} Z),$
$\sigma(t \bar{t} Z + 2~{\rm jets} ) \simeq \alpha_s^2 \sigma(t \bar{t} Z ),$
etc. --- where each extra QCD jet costs a factor of $\alpha_s$ (with
$\alpha_s$ typically evaluated at the relevant $p_T$ scale of
$\simeq 40~{\rm GeV}).$ This simple ansatz is motivated
by the results of Behrends {\it et al.}\cite{behrends}
who have found, when calculating $W$ plus jets cross sections (including
realistic acceptance and isolation criteria), that
$ R_n = \sigma( W + n~{\rm jets} ) / \sigma( W + (n-1)~{\rm jets} ) \simeq .2 $
at Tevatron energies. Though this procedure certainly has its limitations
and has not yet been demonstrated to hold at SSC energies and above
(with a corresponding
small value of $R_n$), we will nevertheless adopt it as a rough estimate
and use a factor of $(.2)^n$ to dress cross sections of Table~2 with
$n$ additional QCD jets. The denominator of each entry of Tables~3,4
is a sum of the contribution in parenthesis and all the
relevant ``QCD dressed'' contributions.
For fixed $n_{Z \rightarrow \ell \bar{\ell} }, n_{W \rightarrow \ell \nu},$
the ``QCD dressed'' contribution to the $n_{\rm jets}$ background
is equal to $(.2)^2$ times the total background for $n_{\rm jets}-2.$

Diagonal entries running from the lower left to upper right of
Tables~3,4 correspond to signals with a fixed
minimum value of $n_w+n_z.$
Consider, for example, the entry for
($n_{W \rightarrow \ell \nu} = 5, n_{\rm jet} = 8)$
in Table~3. As far as the signal is
concerned, the five leptons could only have come from five
leptonic $w$ decays (remember --- we assume 100\% efficiency
in identifying and removing $Z$ decays to $e$ or $\mu$ pairs).
Similarly, the eight jets presumably come from
the hadronic decays of four bosons. Thus the visible
decay products correspond to a minimum number of nine
Goldstone bosons. Only a minimum is determined because there could, in
addition, be an arbitrary number of undetected
$Z \rightarrow \nu \bar{\nu}$ decays.
The same conclusion follows by considering any entry along
the same diagonal.

Table~3, which corresponds to signatures with no leptonically
reconstructed $Z$ decays $(n_{Z \rightarrow \nu \bar{\nu}}=0)$,
indicates that the signal/background ($S/B$) exceeds unity essentially
only for $n_w+n_z \ge 9.$ Summing over signatures of Table~3 which
have  $S/B > 1 $ gives an overall $S/B = 3.0 ~{\rm fb}/0.8~{\rm fb}.$
A detailed decomposition of the background into specific
contributions reveals that when the signal emerges from the
background (along the diagonal corresponding to $n_w+n_z \ge 9$)
the dominant backgrounds are $6t$ production and $4t+{\rm jets}$.

Turning to Table~4, which corresponds to signatures containing
exactly one leptonically reconstructed $Z$ decay
$(n_{Z \rightarrow \nu \bar{\nu}}=1)$,
we find that $S/B > 1$ for signals with
$n_w+n_z \ge 7.$ Summing over all signatures of Table~4 which
have $S/B > 1$ gives an overall $S/B = 3.6~{\rm fb}/0.4~{\rm fb}.$
When the signal emerges from the background (along the diagonal
corresponding to $n_w+n_z \ge 7$ )
the dominant backgrounds are $Ht\bar{t}+ {\rm jets}$
(where $H \rightarrow ZZ$)
and $Zt\bar{t} + {\rm jets}.$

Once again we should emphasize the tentative nature of our numerical
results. They are only meant as a rough indication of whether or not
inelastic multi-$w$ would be observable at a hadron collider.
On the positive side, without exploiting any of the
special signal characteristics
({\it e.g.,} the longitudinal nature of the gauge bosons,
large summed transverse energy, the possibility of reconstructing
hadronic W and Z decays, etc.) we see that observing inelastic $w$ production
is not ruled out (at least for the hypothetical scaled up QCD model
we have used). On the other hand, cross sections of $O(1~{\rm fb})$
with $S/B \simeq O(1)$ are marginal: a realistic model and
a definitive background study could easily introduce cumulative
factors of two which could completely alter (for better or worse)
the prospects of observing a signal. For example, had we not
artificially assumed that all hadronic jets are contained,
isolated,  and identifiable, then the signal would have been
diluted by being spread over signatures with
both even and odd numbers of observed jets.
With qualifications of this type in
mind, our results should only be viewed as a first step towards
demonstrating the plausibility of observing inelastic multi-$w$ production.

\section{Summary and Conclusions}
\label{se:summary}
\hspace*{\parindent}
Given the number and nature of the assumptions we have made,
a brief summary of our whole analysis is in order.
After outlining our scaling procedure,
we found that a $ww$ total cross section of
$\hat{\sigma}_0 = 1~{\rm nb}$ above a $ww$ c.m.s. threshold
$\sqrt{\hat{s}_0} = 5~{\rm TeV}$ corresponds to a $pp$ cross section
of $\simeq 190~{\rm fb}$ at $\sqrt{s} = 200~{\rm TeV}.$
This sets an upper limit on the signal (within the assumptions)
--- before any kind of backgrounds, acceptance, efficiencies,
etc. are considered.

Going further, we scaled up an assumed
hadronic multiplicity distribution and determined how
$\sigma^{pp}_{ {\rm multi-}w } \simeq 190~{\rm fb}$ is partitioned
into contributions with fixed multiplicity (see Fig.~2).
Due to the rapidly falling $ww$ luminosity, most $ww$ interactions
occur just above the assumed threshold $\sqrt{\hat{s}_0}$ which
leads to $\langle n_w + n_z \rangle \simeq 4-5.$ When
backgrounds are considered, a naive analysis reveals that
signals containing $\lwig~6$ Goldstone bosons are likely dominated
by generic backgrounds and suggests that the potentially detectable
signal resides largely in the high-multiplicity tail of the
multiplicity distribution. This restriction reduces the original
$190~{\rm fb}$ approximately by a factor of two.

Signatures roughly consisting of a) at least two high $p_T$
leptons (presumably from $W$ decay) and ten or more jets
or b) one leptonically reconstructed $Z$, one or more
high $p_T$ leptons and eight or more jets
have a combined cross section of a few fb which is a factor
of 4-5 above the background.
Of course, given the nature of the assumptions involved,
these cross sections are not
to be taken literally --- they are only meant to be indicative
of the strengths or weaknesses of a multi-$w$ signal.
A relatively interesting conclusion, however, is that if strong
multi-$w$ production is to be observed over conventional
backgrounds, the inelastic production of three or four
longitudinal bosons is likely not sufficient; one has
to consider the productions of seven or more bosons. The frequency
of such high multiplicity states will likely depend critically upon
the details of the underlying strongly interacting theory.

If one is willing to entertain the notion of
of inelastic $ww$ physics above
a $ww$ c.m.s. threshold as low as $\sqrt{\hat{s}_0} = 1~{\rm TeV},$
then most of our results still hold and are of relevance to the SSC
physics program. As can be deduced from Figure~2,
the total signal rates for
($\sqrt{s}=200~{\rm TeV}, \sqrt{\hat{s}_0} = 5~{\rm TeV}$)
are identical to those for
($\sqrt{s}=40~{\rm TeV}, \sqrt{\hat{s}_0} = 1~{\rm TeV}$).
Since the average charge multiplicity $\langle n_w \rangle$ is
a slowly varying function of $\sqrt{\hat{s}}$ a reduced
threshold does not change the overall $w$ and $z$ multiplicity
significantly. On the other hand, all the background rates at the SSC
are lower than at $\sqrt{s}~=~200~{\rm TeV}.$

The broad conclusion to be drawn is that it appears
feasible to observe multi-$w$ production
arising from a strongly interacting Higgs sector (if such
a sector exists at all). Since the energy scale at which
inelasticity sets in is generally determined by the masses of the
low-lying resonances of the underlying strongly interacting theory,
our results are only suggestive (since they arise from
conservatively assuming a rather high threshold of ${\rm 5~TeV}).$
Nevertheless,
our results are rather encouraging and suggest that the
inelastic $W_L$ and $Z_L$ production may provide an interesting
window to the mechanism of strongly broken electroweak symmetry.

\section{Acknowledgements}
\hspace*{\parindent}
D.M. would like to acknowledge a helpful conversation with A.~Stange.
D.M. is supported by the ELOISATRON Project, R.D.P. is supported in
part by the Department of Energy under Contract DE-AT03-88ER 40384 Mod
A006-Task~C and  R.R. is supported by the Texas National Research
Laboratory Commission under grant number RGFY9114.

\newpage
\section*{Figure Captions}

\begin{description}
\item[Figure 1.] Proton-proton cross section at center of
mass energy $ \sqrt{s} $ for subprocess
cross section $ \hat{\sigma}( w w \rightarrow X ) = \hat{\sigma}_0
\Theta ( \hat{s} - \hat{s}_0  ). $
Arrows indicate situations relevant to the LHC
($\sqrt{s} = 15.4~{\rm TeV}),$
SSC ($\sqrt{s} = 40~{\rm TeV}),$
and ELOISATRON ($\sqrt{s} = 200~{\rm TeV}$) for an assumed threshold
of $\sqrt{\hat{s}_0} = 5~{\rm TeV}.$

\item[Figure 2.] Goldstone boson multiplicity distribution
for $\sqrt{s} = 200~{\rm TeV}$ proton-proton collisions assuming
$\sqrt{\hat{s}_0} = 5~{\rm TeV}.$ The distribution is based on
charged particle multiplicities measured in $e^+e^-$ annihilation
as described in the text.

\item[Figure 3.] Transverse momentum distribution
for final state Goldstone bosons for the subprocess
$w w \rightarrow 8~w$ at $\sqrt{s} = 200~{\rm TeV}$ with
$\sqrt{\hat{s}_0} = ~5~{\rm TeV}$ for
$p_T$ scale $\langle p_T \rangle_{w} = 80~{\rm GeV}$ (solid) and
$\langle p_T \rangle_{w} = 1~{\rm TeV}$ (dashed).

\item[Figure 4.] Laboratory rapidity distribution for
final state Goldstone bosons for the subprocess
$w w \rightarrow 8~w$ at $\sqrt{s} = 200~{\rm TeV}$ with
$\sqrt{\hat{s}_0} = ~5~{\rm TeV}$ for
$p_T$ scale $\langle p_T \rangle_{w} = 80~{\rm GeV}$ (solid) and
$\langle p_T \rangle_{w} = 1~{\rm TeV}$ (dashed).
\end{description}

\newpage

\begin{center}
\begin{description}
\item[Table 1.]
Effective decay branching fractions used to estimate the
signal and background cross sections for jet plus lepton
signatures. Notes: (1)
summed over $e$ and $\mu$; (2) as a simplification,
$\tau$ decays counted as two-jet decays in this context;
(3) assumes $m_H = 400~{\rm GeV}$, $m_t = 140~{\rm GeV}$,
$m_W = 80~{\rm GeV}$ and $m_Z = 91.17~{\rm GeV}.$

\end{description}
\end{center}
\begin{center}
 \begin{tabular} { | l l l | } \hline
 & & \\
${\rm Br}(W \rightarrow \ell \nu)$           & = & $2/9^{(1)}$ \\
${\rm Br}(W \rightarrow 2~{\rm jets})$       & = & $7/9^{(2)}$ \\
 & & \\
${\rm Br}( Z \rightarrow \ell \bar{\ell}  )$ & = & $.067^{(1)}$ \\
${\rm Br}( Z \rightarrow \nu  \bar{\nu } ) $ & = & .2           \\
${\rm Br}( Z \rightarrow 2 ~{\rm jets}   ) $ & = & $.733$ \\
 & & \\
${\rm Br}( b \rightarrow \ell + {\rm 1~jet}) $&= &$2/9^{(1)}$ \\
${\rm Br}( b \rightarrow  1~{\rm jet}      ) $&= &$7/9$ \\
 & & \\
${\rm Br}( t \rightarrow W b )$ & = & 1  \\
 & & \\
${\rm Br}( H \rightarrow W^+ W^-   )$ & = & $.55^{(3)}$ \\
${\rm Br}( H \rightarrow Z   Z     )$ & = & .26        \\
${\rm Br}( H \rightarrow t \bar{t} )$ & = & .19        \\
& & \\
  \hline
 \end{tabular}
\end{center}

\newpage
\begin{center}
\begin{description}
\item[Table 2.] Standard Model
backgrounds ($m_t = 140~{\rm GeV},$ $m_H = 400~{\rm GeV}$)
for multiple gauge boson production
at $\sqrt{s} = 200~{\rm TeV}$
obtained by extrapolating the results of Ref.~\cite{bargerweakbosonpaper}
except where noted.
(1) Nonresonant contributions taken from Ref.~\cite{bargerfourweakboson}.
(2) Estimates obtained by assuming $\sigma(4Z) = f^4 \sigma(4W)$
so that each additional $Z$ boson suppresses $\sigma(4W)$ by
a factor $f \simeq .44.$
(3) Estimates obtained by extrapolating LHC and SSC rates of
Ref.~\cite{bargerweakbosonpaper}
assuming a linear relationship between $\ln \sigma$
and $\ln s.$
(4) Estimate obtained by assuming $\sigma(6t) / \sigma(4 t)
= \sigma(4t) / \sigma (2 t )$.

\end{description}
\end{center}
\begin{center}
\begin{tabular} {| l | r | l | r |} \hline
  & & & \\
 Process                                     & $ \sigma$ ~ (fb) &
 Process                                     & $ \sigma$ ~ (fb) \\
  & & & \\ \hline
  & & & \\

$t \bar{t}                $                   &  $ 1.1 \times 10^8  $  &
$ t \bar{t} t \bar{t}     $                   &  $ 18000 $  \\

$ t \bar{t} b \bar{b}     $                   &  $ 1.6\times10^6 $ &
$ t \bar{t} H             $                   &  $  8600 $  \\

$g g \rightarrow H        $                   &  $ 2.6 \times 10^5$  &
$ t \bar{t} W^+ W^-       $                   &  $  3100 $  \\

$W^+ Z + W^- Z            $                   &  $ 2.7 \times 10^5 $ &
$ t \bar{t} Z Z           $                   &  $   380 $  \\

$W^+ W^-                  $                   &  $ 2.4 \times 10^5 $ &
$ t \bar{t}W^+ Z  +  t\bar{t}W^-Z$            &  $   130 $  \\

$Z   Z                    $                   &  $ 1.0 \times 10^5 $ &
$ H H                     $                   &  $   100 $  \\

$q q \rightarrow q q H    $                   &  $ 3.1 \times 10^4$ &
$ W^+ W^- W^+ W^-         $                   &  $    50^{[1]}$  \\

$ q q \rightarrow q q W^+ W^+ $               &  $ 1.8 \times 10^4 $ &
$ WWWZ                    $                   &  $    22^{[2]}$  \\

$ q q \rightarrow q q W^- W^- $               &  $ 1.2 \times 10^4 $ &
$ WWZZ                    $                   &  $    10^{[2]}$  \\

                                              &                  &
$ WZZZ                    $                   &  $     4^{[2]}$  \\

$ t \bar{t} Z             $                   &  $ 120000$ &
$ ZZZZ                    $                   &  $     2^{[1]}$  \\

$ t \bar{t} W^+ + t \bar{t} W^- $             &  $  9300 $ &
                                              &             \\

$ W^+ W^+ W^- + W^- W^- W^+     $             &  $  2400 $ &
                                              &             \\

$ W^+ W^- Z               $                   &  $  2000 $ &
$ HHW^+ + HHW^-           $                   &  $     2^{[3]}$  \\

$ W^+ Z Z  + W^- Z Z      $                   &  $   560 $ &
$ HHZ                     $                   &  $     1^{[3]}$  \\

$ W^+ H + W^- H           $                   &  $   260 $ &
 & \\

$ Z H                     $                   &  $   180 $ &
   & \\

$ Z Z Z                   $                   &  $   150 $ &
$ t \bar{t} t \bar{t} t \bar{t}  $            &  $      3^{[4]}$   \\
 & & & \\
\hline
\end{tabular}
\end{center}

\newpage
\begin{center}
\begin{description}
\item[Table 3.]
Signal/Background cross sections (in fb) at $\sqrt{s} = 200~{\rm TeV}$
for signatures with no
leptonic $Z$ decay ($n_{Z \rightarrow \ell \bar{\ell} } = 0 $).
First quantity in denominator is total background; quantity in parenthesis is
contribution to total from processes of Table~2.
Cross sections less than $.005$~fb are rounded to zero.

\end{description}
\end{center}
\setlength{\unitlength}{1cm}
\begin{picture}(10,11)(2.6,0.5)

\put(0,6){
 \begin{tabular}
  { | c |
    c |
    c |
    c |
    c |
    c |
    c |
    c | } \hline
  &
  &
  &
  &
  &
  &
  &
  \\
  $n_{W \rightarrow \ell \nu} ~ \backslash ~ n_{\rm jet}$ &
           6 &
           8 &
          10 &
          12 &
          14 &
          16 &
          18 \\
  &
  &
  &
  &
  &
  &
  &
  \\
  \hline
  &
  &
  &
  &
  &
  &
  &
  \\
           2 &
 $ \begin{array}{c}
\displaystyle{           4\over       30000(9000)}  \\
 \end{array} $ &
 $ \begin{array}{c}
\displaystyle{           4\over        3000(1000)}  \\
 \end{array} $ &
 $ \begin{array}{c}
\displaystyle{           3\over         200(90)} \\
 \end{array} $ &
 $ \begin{array}{c}
\displaystyle{           2\over           8(.8)}  \\
 \end{array} $ &
 $ \begin{array}{c}
\displaystyle{           1\over           .5(.2)}  \\
 \end{array} $ &
 $ \begin{array}{c}
\displaystyle{           .5\over           .03(.01)} \\
 \end{array} $ &
 $ \begin{array}{c}
\displaystyle{           .2\over           0} \\
 \end{array} $ \\
  &
  &
  &
  &
  &
  &
  &
  \\
  \hline
  &
  &
  &
  &
  &
  &
  &
  \\
           3 &
 $ \begin{array}{c}
\displaystyle{           1\over         500(300)} \\
 \end{array} $ &
 $ \begin{array}{c}
\displaystyle{           1\over          60(40)} \\
 \end{array} $ &
 $ \begin{array}{c}
\displaystyle{           .7\over           3(.9)}  \\
 \end{array} $ &
 $ \begin{array}{c}
\displaystyle{           .5\over           .2(.07)} \\
 \end{array} $ &
 $ \begin{array}{c}
\displaystyle{           .3\over           .02(.01)}  \\
 \end{array} $ &
 $ \begin{array}{c}
\displaystyle{           .1\over           0} \\
 \end{array} $ &
 $ \begin{array}{c}
\displaystyle{           .06\over           0} \\
 \end{array} $ \\
  &
  &
  &
  &
  &
  &
  &
  \\
  \hline
  &
  &
  &
  &
  &
  &
  &
  \\
           4 &
 $ \begin{array}{c}
\displaystyle{           .2\over           9(7)} \\
 \end{array} $ &
 $ \begin{array}{c}
\displaystyle{           .2\over           .8(.4)} \\
 \end{array} $ &
 $ \begin{array}{c}
\displaystyle{           .1\over           .05(.02)} \\
 \end{array} $ &
 $ \begin{array}{c}
\displaystyle{           .09\over           .01(0)} \\
 \end{array} $ &
 $ \begin{array}{c}
\displaystyle{           .05\over           0} \\
 \end{array} $ &
 $ \begin{array}{c}
\displaystyle{           .02\over           0} \\
 \end{array} $ &
 $ \begin{array}{c}
\displaystyle{           .01\over           0} \\
 \end{array} $ \\
  &
  &
  &
  &
  &
  &
  &
  \\
  \hline
  &
  &
  &
  &
  &
  &
  &
  \\
           5 &
 $ \begin{array}{c}
\displaystyle{           .03\over           .10(.08)} \\
 \end{array} $ &
 $ \begin{array}{c}
\displaystyle{           .02\over           .01(0)} \\
 \end{array} $ &
 $ \begin{array}{c}
\displaystyle{           .02\over           0} \\
 \end{array} $ &
 $ \begin{array}{c}
\displaystyle{           .01\over           0} \\
 \end{array} $ &
 $ \begin{array}{c}
\displaystyle{           .01\over           0} \\
 \end{array} $ &
 $ \begin{array}{c}
\displaystyle{           0\over           0} \\
 \end{array} $ &
 $ \begin{array}{c}
\displaystyle{           0\over           0} \\
 \end{array} $ \\
  &
  &
  &
  &
  &
  &
  &
  \\
  \hline
 \end{tabular} }
\end{picture}
\newpage
\oddsidemargin=.5in
\begin{center}
\begin{description}
\item[Table 4]
Signal/Background cross sections (in fb) at $\sqrt{s} = 200~{\rm TeV}$
for signatures with one leptonic $Z$ decay
 ($n_{Z \rightarrow \ell \bar{\ell} } = 1 $).
First quantity in denominator is total background; quantity in parenthesis is
contribution to total from processes of Table~2.
Cross sections less than $.005$~fb are rounded to zero.
\end{description}
\end{center}
\begin{center}
 \begin{tabular}
  { | c |
    c |
    c |
    c |
    c |
    c |
    c |
    c | } \hline
  &
  &
  &
  &
  &
  &
  &
  \\
  $n_{W \rightarrow \ell \nu}  ~ \backslash ~ n_{\rm jet}$ &
           6 &
           8 &
          10 &
          12 &
          14 &
          16 &
          18 \\
  &
  &
  &
  &
  &
  &
  &
  \\
  \hline
  &
  &
  &
  &
  &
  &
  &
  \\
           1 &
 $ \begin{array}{c}
\displaystyle{           1\over         200(100)} \\
 \end{array} $ &
 $ \begin{array}{c}
\displaystyle{           1\over           8(1)}  \\
 \end{array} $ &
 $ \begin{array}{c}
\displaystyle{           .9\over           .3(0)} \\
 \end{array} $ &
 $ \begin{array}{c}
\displaystyle{           .6\over           .01(0)} \\
 \end{array} $ &
 $ \begin{array}{c}
\displaystyle{           .3\over           0} \\
 \end{array} $ &
 $ \begin{array}{c}
\displaystyle{           .2\over           0} \\
 \end{array} $ &
 $ \begin{array}{c}
\displaystyle{           .07\over           0} \\
 \end{array} $ \\
  &
  &
  &
  &
  &
  &
  &
  \\
  \hline
  &
  &
  &
  &
  &
  &
  &
  \\
           2 &
 $ \begin{array}{c}
\displaystyle{           .5\over           3(.9)} \\
 \end{array} $ &
 $ \begin{array}{c}
\displaystyle{           .4\over           .1(.01)}  \\
 \end{array} $ &
 $ \begin{array}{c}
\displaystyle{           .3\over           0} \\
 \end{array} $ &
 $ \begin{array}{c}
\displaystyle{           .2\over           0} \\
 \end{array} $ &
 $ \begin{array}{c}
\displaystyle{           .1\over           0} \\
 \end{array} $ &
 $ \begin{array}{c}
\displaystyle{           .06\over           0} \\
 \end{array} $ &
 $ \begin{array}{c}
\displaystyle{           .03\over           0} \\
 \end{array} $ \\
  &
  &
  &
  &
  &
  &
  &
  \\
  \hline
  &
  &
  &
  &
  &
  &
  &
  \\
           3 &
 $ \begin{array}{c}
\displaystyle{           .1\over           .02(0)}  \\
\end{array} $ &
 $ \begin{array}{c}
\displaystyle{           .1\over           0} \\
 \end{array} $ &
 $ \begin{array}{c}
\displaystyle{           .08\over           0} \\
 \end{array} $ &
 $ \begin{array}{c}
\displaystyle{           .05\over           0} \\
 \end{array} $ &
 $ \begin{array}{c}
\displaystyle{           .03\over           0} \\
 \end{array} $ &
 $ \begin{array}{c}
\displaystyle{           .02\over           0} \\
 \end{array} $ &
 $ \begin{array}{c}
\displaystyle{           .01\over           0} \\
 \end{array} $ \\
  &
  &
  &
  &
  &
  &
  &
  \\
  \hline
  &
  &
  &
  &
  &
  &
  &
  \\
           4 &
 $ \begin{array}{c}
\displaystyle{           .02\over           0} \\
 \end{array} $ &
 $ \begin{array}{c}
\displaystyle{           .02\over           0} \\
 \end{array} $ &
 $ \begin{array}{c}
\displaystyle{           .02\over           0} \\
 \end{array} $ &
 $ \begin{array}{c}
\displaystyle{           .01\over           0} \\
 \end{array} $ &
 $ \begin{array}{c}
\displaystyle{           .01\over           0} \\
 \end{array} $ &
 $ \begin{array}{c}
\displaystyle{           0\over           0} \\
 \end{array} $ &
 $ \begin{array}{c}
\displaystyle{           0\over           0} \\
 \end{array} $ \\
  &
  &
  &
  &
  &
  &
  &
  \\
  \hline
 \end{tabular}
\end{center}
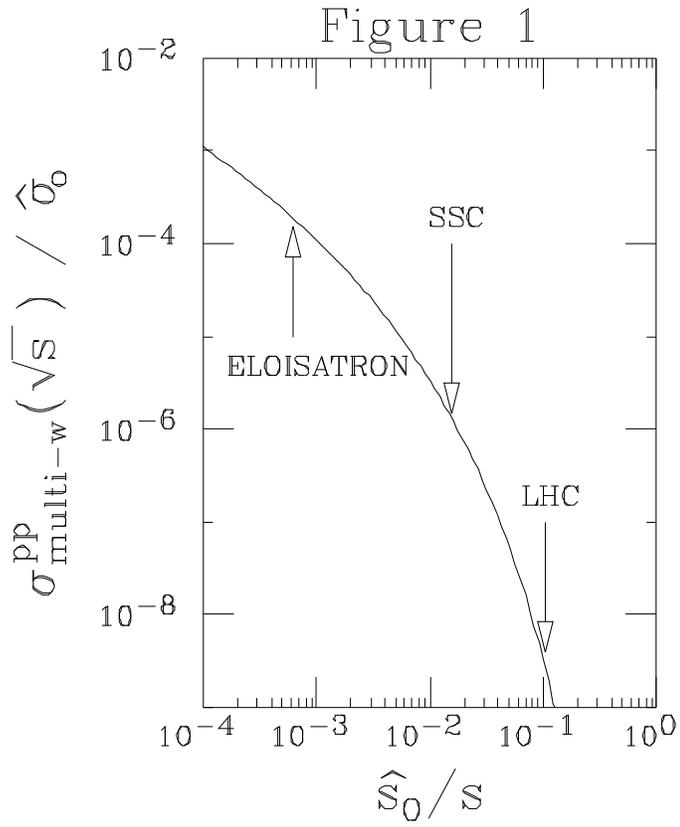
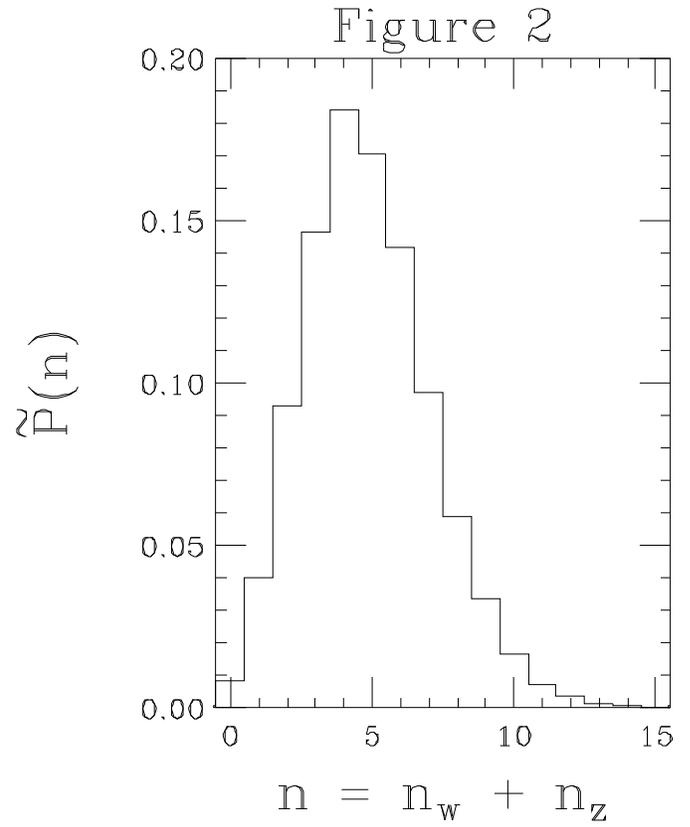
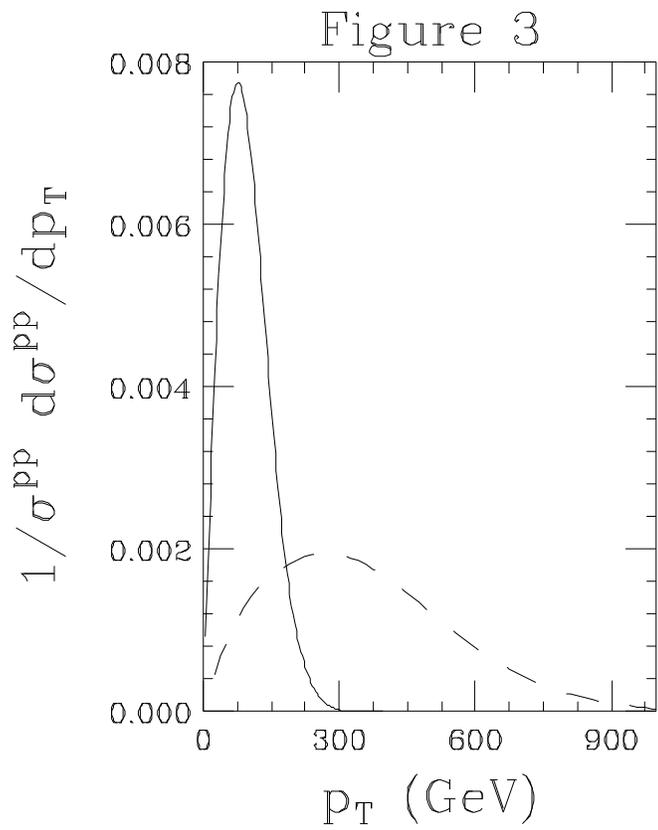
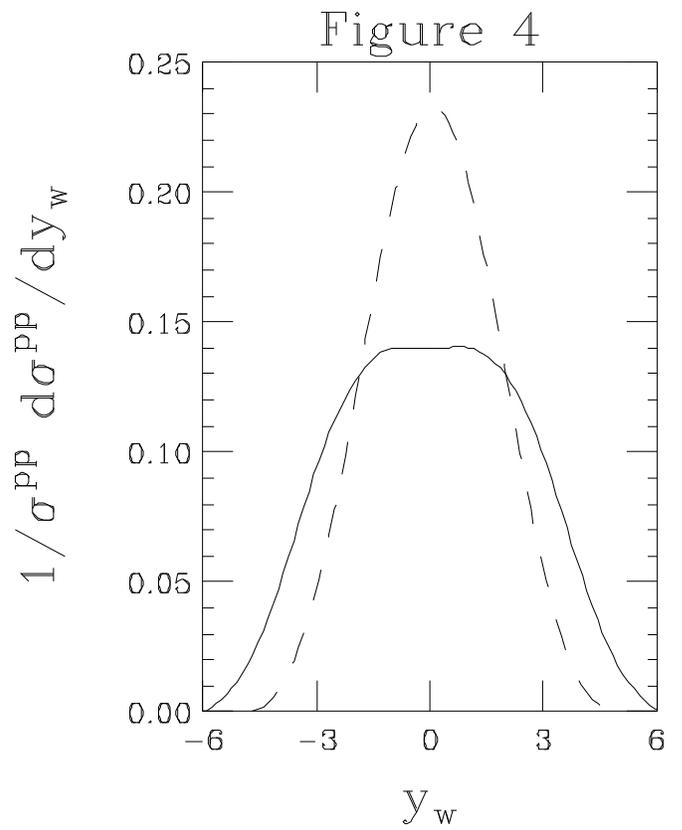
\end{document}